\documentclass[prl, aps, amssymb, twocolumn, superscriptaddress, notitlepage, longbibliography]{revtex4-2}

\usepackage{amsmath}
\usepackage{amsfonts}
\usepackage{graphicx} 
\usepackage{float}
\usepackage{braket}
\usepackage{xcolor}
\usepackage{subfigure}
\usepackage{cancel}
\usepackage[colorlinks,linkcolor=blue,citecolor=blue,urlcolor=blue]{hyperref}

\begin{document}

\title{Floquet-Anderson localization in the Thouless pump and how to avoid it}

\author{Andr\'as Grabarits}
\affiliation{Department of Theoretical Physics, Institute of Physics, 
Budapest University of Technology and Economics, M\H uegyetem rkp. 3., H-1111 Budapest, Hungary}
\affiliation{MTA-BME Quantum Dynamics and Correlations Research Group, 
Budapest University of Technology and Economics,  M\H uegyetem rkp. 3., H-1111 Budapest, Hungary}
\affiliation{Department of Physics and Materials Science, University of Luxembourg, L-1511 Luxembourg, Luxembourg}
\author{Attila Tak\'acs}
\affiliation{Department of Theoretical Physics, Institute of Physics, 
Budapest University of Technology and Economics, M\H uegyetem rkp. 3., H-1111 Budapest, Hungary}
\affiliation{Université de Lorraine, CNRS, LPCT, F-54000 Nancy, France}
\author{Ion Cosma Fulga}
\affiliation{Leibniz Institute for Solid State and Materials Research,
IFW Dresden, Helmholtzstrasse 20, 01069 Dresden, Germany}
\affiliation{W\"{u}rzburg-Dresden Cluster of Excellence ct.qmat, 01062 Dresden, Germany}

\author{J\'anos K. Asb\'oth}
\affiliation{Department of Theoretical Physics, Institute of Physics, 
Budapest University of Technology and Economics, M\H uegyetem rkp. 3., H-1111 Budapest, Hungary}
\affiliation{Wigner Research Centre for Physics, H-1525 Budapest, P.O. Box 49., Hungary}

\begin{abstract}
We investigate numerically how onsite disorder affects conduction in the  periodically driven Rice-Mele model, a prototypical realization of the Thouless pump.  
Although the pump is robust against disorder in the fully adiabatic limit, much less is known about the case of finite period time $T$, which is relevant also in light of recent experimental realizations. 
We find that at any fixed period time and nonzero disorder, increasing the system size $L\to\infty$ always leads to a breakdown of the pump, indicating Anderson localization of the Floquet states. 
Our numerics indicate, however, that in a properly defined thermodynamic limit, where $L/T^\theta$ is kept constant, Anderson localization can be avoided, and the charge pumped per cycle has a well-defined value -- as long as the disorder is not too strong.   
The critical exponent $\theta$ is not universal, rather, its value depends on the disorder strength. 
Our findings 
are relevant for 
practical, experimental realizations of the Thouless pump, 
for studies investigating the nature of its current-carrying Floquet eigenstates, 
as well as the mechanism of the full breakdown of the pump, expected if the disorder exceeds a critical value. 
\end{abstract}

\maketitle

The Thouless pump \cite{thouless1983quantization} was instrumental in understanding the role topology plays in the theory of the quantum Hall effect. 
Its simplest form is that of a two-band, gapped fermionic chain
whose parameters are slowly and periodically varied in time. 
In the half-filled state, where the lower/upper band is completely filled/empy, and in the adiabatic limit, an integer number $Q$ of fermions are pumped in the lower band, where $Q$ is a topological invariant, equal to the Chern number $C$ of the pump sequence. 

While it started out as a thought experiment, the Thouless pump can now be found in the lab \cite{citro2023thouless}.
After its demonstration in photonic systems \cite{Kraus2012, Verbin2015, Ke2016, Cerjan2020}, topological pumping has been realized in a variety of platforms, such as mechanical metamaterials \cite{Grinberg2020, Xia2021},
ultracold atoms \cite{lohse2016thouless, nakajima2016topological, lu2016geometrical}, and other quantum systems \cite{Gong2018}.
More recently, an electrical circuit demonstration 
has been put forward \cite{Stegmaier2023}.

The real-life Thouless pump has a finite size, $L$, a finite period time, $T$, and it is disordered.
This raises the question whether these, either separately or in combination, can prove detrimental to its robustness, namely to the quantization of the pumped charge.
For instance, even without disorder, a finite period time gives corrections to the quantized value of the pumped charge. For a quench-like switching on of the pump, Ref.~\cite{Privitera2018} found these to 
scale as $1/T^2$, but they should be greatly reduced for a smoother switching-on of the periodic driving cycle.
The breaking of adiabaticity, however, is not always sufficient to destroy the Thouless pump.
As shown in Ref.~\cite{Lychkovskiy2017}, if the disorder-free pump is made longer and longer, then adiabaticity will be strongly broken for any finite $T$, no matter how large.
In spite of this, they found that the quantization of the pumped charge survives in the steady-state regime, when the pump performs cycle after cycle. 
Similarly, in the adiabatic $T\to\infty$ limit,  
the quantization of the pumped charge should be robust against small disorder, as discussed by Thouless and Niu \cite{niu1984quantised}. 
Here, eigenstates of the time-dependent Hamiltonian are all Anderson localized, and in the adiabatic limit an intuitive (although possibly misleading) picture is that periodic modulation pumps charge between them.

In this Letter we focus on the effect of onsite disorder on an actual Thouless pump (finite $T$ rather than the adiabatic limit).
On the one hand, disorder 
can even result in a suppression of finite-$T$ corrections, and a higher pumped charge \cite{wauters2019localization}. 
On the other, adding too much disorder has to result in a breakdown of the pump, via an Anderson localization transition -- a few works have already studied this numerically \cite{wauters2019localization, Hayward2021}. 

Onsite disorder on the Thouless pump is particularly interesting because of the connection to the ``levitation and annihilation'' 
in Chern insulators \cite{Laughlin1984}. 
Disorder in the Chern insulator localizes its eigenstates, but each topological band has (at least) one state that remains extended in the thermodynamic limit, which ``carries the Chern number'' \cite{Halperin1982, Thouless1984, Thonhauser2006}. 
As disorder is increased, full Anderson localization happens by these robustly extended states ``levitating'' towards each other in energy and ``annihilating''. 
Can such phenomena be observed in the Floquet states 
of the disordered Thouless pump? 
Numerical results \cite{wauters2019localization} are consistent with this, and have even identified a critical exponent for this Anderson localization transition, obtained by scaling up the size of the pump at a constant (and large) period time. 

There is an issue with the disordered Thouless pump in the thermodynamic limit, however, that to the best of our knowledge has not been directly addressed yet.
One might think that for a thermodynamic limit, $L\to\infty$ and $T\to\infty$ should be taken one after the other. However, we argue this is inadequate. 
Taking $T\to \infty$ first is problematic, since the charge pump becomes infinitely slow. 
Moreover, in this ``ultra-adiabatic'' limit transitions occur between distant Anderson localized eigenstates, thus computation of $Q$ needs open boundary conditions \cite{Imura2018}.   
Taking $L\to\infty$ first is often (sometimes tacitly) assumed. 
However, as we show in the following, this leads to a breakdown of the pump due to the Anderson localization of the Floquet eigenstates. 
In this Letter we suggest a properly defined way to take $L\to\infty$ and $T\to\infty$.

\emph{The model.} ---
We consider 
the periodically driven Rice-Mele model \cite{rice1982elementary} with an onsite potential disorder that is independent of time. 
Spinless fermions hop on a closed chain of $L=2N$ sites, with 
the unitary time evolution governed by the 
Hamiltonian,
\begin{align}
\hat H(t) &=-\sum_{m=1}^L  \left[ 
J+(-1)^m \tilde{J} \cos\tfrac{2\pi t}{T} \right] 
\hat{c}^\dagger_{m} \hat{c}_{m+1}
+ \mathrm{h. c}. \nonumber \\
 & -\sum_{m=1}^L \left[(-1)^m \Delta \sin\tfrac{2\pi t}{T}
 + W \zeta_m \right] \hat{c}^\dagger_m \hat{c}_m,
\end{align}
where $\hat{c}_m$ annihilates a fermion on site $m$, with 
$\hat{c}_{L+1}=\hat{c}_1$, i.e., periodic boundary conditions, 
$J$/$\tilde{J}$ are uniform/staggered components of the nearest-neighbor hopping, $\Delta$ is a staggered onsite potential, and $t$ and $T$ are time and period time.
The onsite disorder has amplitude $W$ and the $\zeta_m$'s are real random numbers uniformly distributed on $[-1/2,1/2]$. 
We set $\hbar=1$ for convenience.
In this noninteracting model, all quantitites of interest can be computed from the single-particle 
$L \times L$ Hamiltonian matrix 
$H(t)$, 
with
$ \hat{H}(t) = \sum_{l,m=1}^L \hat{c}_l^\dagger H_{lm}(t) \hat{c}_m $. 

We use the basis of Floquet states: 
eigenstates $\ket{\psi_n}$ of the single-particle (Floquet) unitary operator $\hat U$ for one period of time evolution,  
$\hat{U} \ket {\psi_n} = e^{-i\varepsilon_n} \ket{\psi_n}$. 
Here 
$\hat{U} =\mathcal T e^{-i\int_0^T \mathrm d t \hat{H}(t)}$, 
where ${\cal T}$ is time ordering, 
$n=1, 2, \ldots, L$ is the eigenstate index 
and $\varepsilon_n$ is the quasienergy.
Floquet states evolve periodically in time, up to a 
phase factor: 
\begin{align}
\ket{\psi_n(t)} &= \mathcal T\,e^{-i\int_0^t \mathrm d t^\prime 
\hat{H}(t^\prime)} \ket{\psi_n}
=e^{i\varepsilon_n T}\ket{\psi_n(t+T)}. 
\end{align}
If the disorder is weak and the pump is run slowly enough, Floquet states can be assigned to bands according to their average energy,
\begin{align}
\overline{E_n}&= 
\frac{1}{T} \int_0^T \mathrm dt \bra{\psi_n(t)}
\hat{H}(t) \ket{\psi_n(t)}. 
\end{align}
Floquet states carry current, whose integral over the time period gives the pumped charge in that state,  
\begin{align}
\label{eq:Qn}
Q_n &= 2 \int_0^T \mathrm dt 
\left( J+ \tilde{J}
\cos\tfrac{2\pi t}{T} 
\right) 
\mathrm{Im} 
[ \psi_{n, 2}^\ast(t) \psi_{n, 1}(t)].     
\end{align}
Here we take the current between sites 1 and 2, but the position does not matter, due to the 
periodicity of the time evolution of Floquet states. 

We calculate the charge pumped in the so-called sustained pumping limit of a filled lower band \cite{Russomanno2012, wauters2019localization}: 
The system is initialized at $t=0$, with the $L/2$ lowest energy eigenstates $\ket{\phi_l}$ of the instantaneous Hamiltonian fully occupied, and then is time evolved. 
After many cycles,  this results effectively in a Floquet diagonal ensemble \cite{Russomanno2012}, i.e., an incoherent mixture where Floquet states are populated with the same weights as at $t=0$. 
Thus, the charge pumped per cycle in this limit is     
\begin{align}
Q &= \sum_{n =1}^L Q_n \sum_{l=1}^{L/2}  
 \left| {\bra{\phi_l}\psi_n\rangle} \right|^2 .
\end{align}

\emph{The numerical method.} ---
We compute the time evolution of the Floquet states, needed for Eq.~\eqref{eq:Qn}, as a matrix product of 
time-slices of the timestep operator.
For the time-slices, we used a recently-developed method based on the Chebyshev polynomial representation of skew Hermitian matrices, $e^{-i H\mathrm dt} \approx 
\alpha_0 - i z_0 H \mathrm dt
- \alpha_1 [H \mathrm dt]^2 + i z_1 [H \mathrm dt]^3 
+ \alpha_2 [H \mathrm dt]^4 
- i z_2 [H \mathrm dt]^5$,
with constants 
specified to 20 decimals \cite{Bader2022}. 
This gives the matrix exponential to numerical accuracy, as long as  $\lvert\lvert H(t) \mathrm dt \rvert\rvert_1 < 1.17\times10^{-2}$.
We could reach chains lengths up to $L=10000$, more than a factor of 10 larger than previous works \cite{wauters2019localization}.
We use the hopping $J$ as our energy scale, and set parameters as
\begin{align}
    J &= 1; &
    \tilde{J}&= 1/2; & 
    \Delta &= 1.5,
\end{align} 
for a well-defined gap with a Chern number $C=1$. 
Thus, $Q=1$ in the adiabatic limit, as long as the instantaneous Hamiltonian is gapped, i.e., $W \lesssim 3.5$ \cite{wauters2019localization}.

\begin{figure}[h]
\includegraphics[width=\columnwidth,,trim={3cm .2cm 6.3cm 1.5cm},clip]{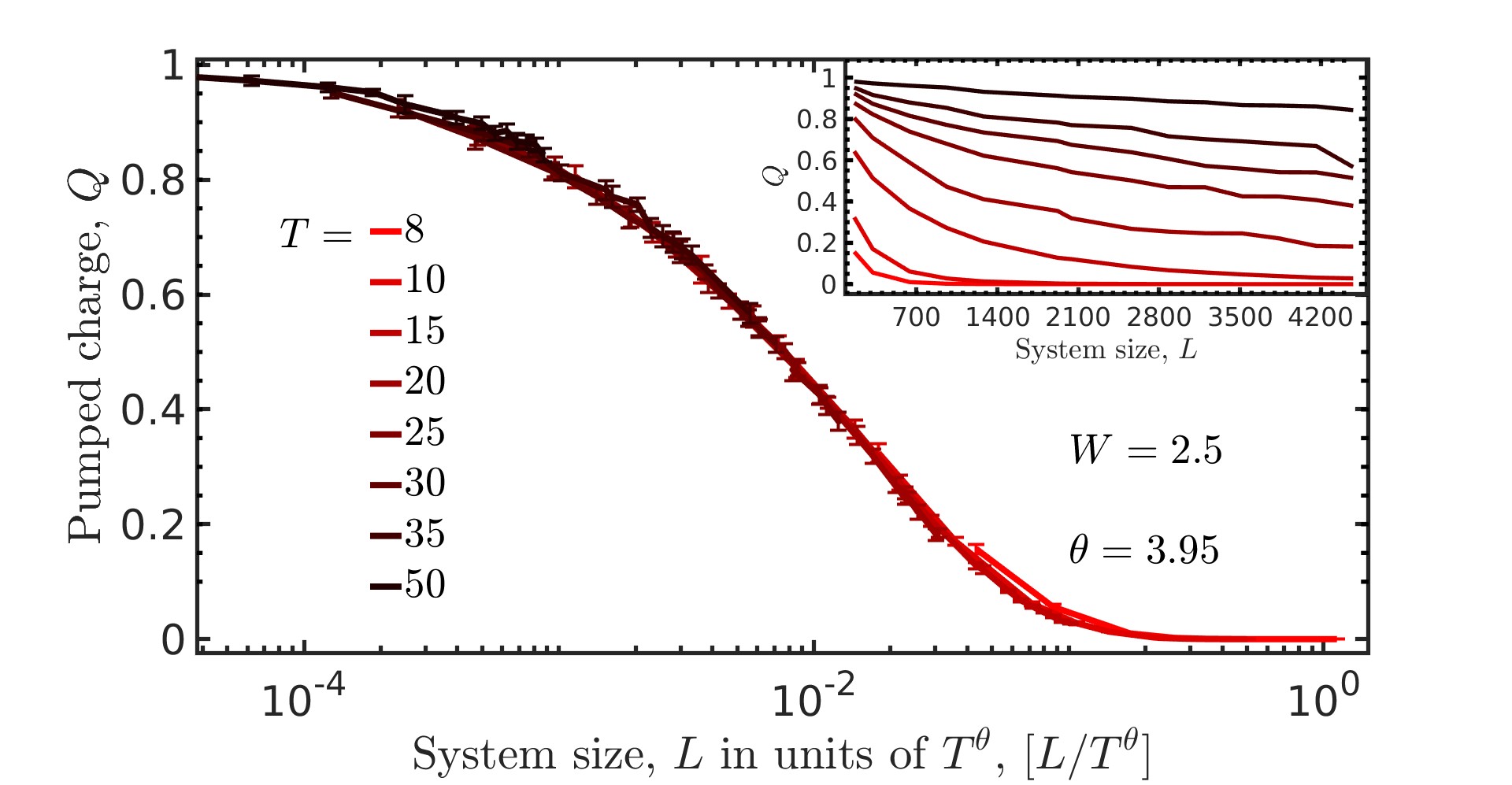}
\caption{
Pumped charge $Q$ decreases as chain length $L$ is increased from $80$ to $4480$, with disorder $W=2.5$ (average of 20 disorder realizations). 
Results for various period times $T$ fall onto each other if rescaling the system size as $L/{T^\theta}$, with $\theta = 3.95$. 
Inset: unscaled data.}
\label{fig:pump_breakdown_length}
\end{figure}

\emph{Results.} --- 
We found that the disordered Thouless pump breaks down when increasing the length $L$ of the chain, keeping the period time $T$ constant. 
Examples are shown in Fig.~\ref{fig:pump_breakdown_length}, for 
disorder $W=2.5$, for $T=8$ to $50$, and $L=80$ up to $4480$
In these and all cases we studied, the pumped charge decreased as the length was increased.
This suggests that Floquet-Anderson localization does not only set in when the disorder is large ($W>3.5$), but occurs for any onsite disorder, $W>0$.
The length $L$ where the pumped charge 
decreases significantly (e.g., $Q=1/2$, or $Q=1/4$) provides an estimate for the 
Floquet localization length $\zeta_F$. 

We found that slower driving (longer period times $T$) leads to more resilient Thouless pumps, with an apparent power-law relation 
\begin{align}
\label{eq:scaling_exponent_def}
    Q(L, T, W) &= Q(L', (L'/L)^{1/\theta(W)} T, W).
\end{align}
This is suggested by the good collapse of the numerically measured $Q$ values when using the above scaling relation, as shown in Fig.~\ref{fig:pump_breakdown_length}. 
Thus, the Floquet localization length appears to scale with the period time as $\zeta_F \propto T^\theta$. 
    	
\begin{figure}
\includegraphics[width=\columnwidth,trim={3.5cm 0.55cm 6.3cm 2.6cm},clip]{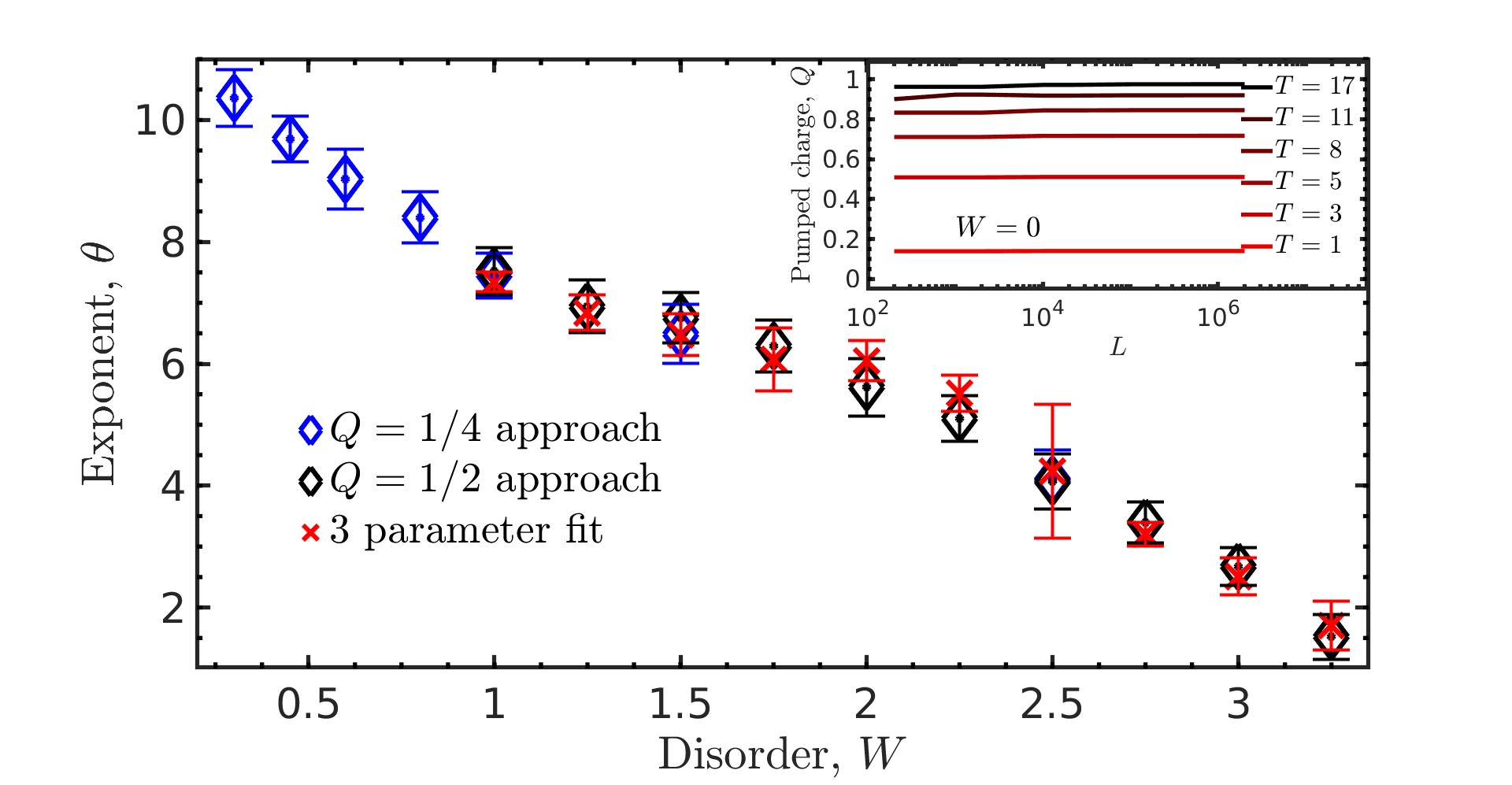}
\caption{
The scaling exponent $\theta$, obtained by  
three numerical approaches for each value of disorder $W$ (see main text for details).
The exponent decreases as $W$ increases,  
while it appears to diverge at $W=0$, consistent with the pumped charge $Q$ being independent of the system size in the clean case -- this latter is shown in the inset.
For a detailed description of how the exponent was extracted see the Supplemental Material \cite{SM}.}
\label{fig:exponent_disorder}
\end{figure}
     
We found that the exponent $\theta$ of Eq.~\eqref{eq:scaling_exponent_def}
does not take on a universal value, but depends continuously on the disorder $W$, as shown in Fig.~\ref{fig:exponent_disorder}. We extracted the exponent 
by three different methods. 
First and second, by identifying $\zeta_F$ with the system size where the pumped charge is $Q=1/2$, and $Q=1/4$, respectively. 
Third, we took all the data for a fixed $W$, and various $T$ and $L$ values, and fitted it with a three-parameter Ansatz - detailed in the Supplemental Material (SM, \cite{SM}). 
These methods agree, and give a disorder-dependent exponent $\theta$, which approaches $\theta\approx 2$ near the critical disorder $W\approx 3.5$. 
Note, however, that the numerical evidence for the power-law scaling is strong only for the case of moderate disorder. 
For smaller disorders, $W \lesssim 1.5$, we have $1/\theta \ll 1$, thus in the numerically available range the evidence for the power-law scaling here is not conclusive, 
as discussed in the SM \cite{SM}.

\begin{figure}[tb]
\includegraphics[width=\columnwidth,trim={1.7cm .55cm 1.5cm 1.7cm},clip]{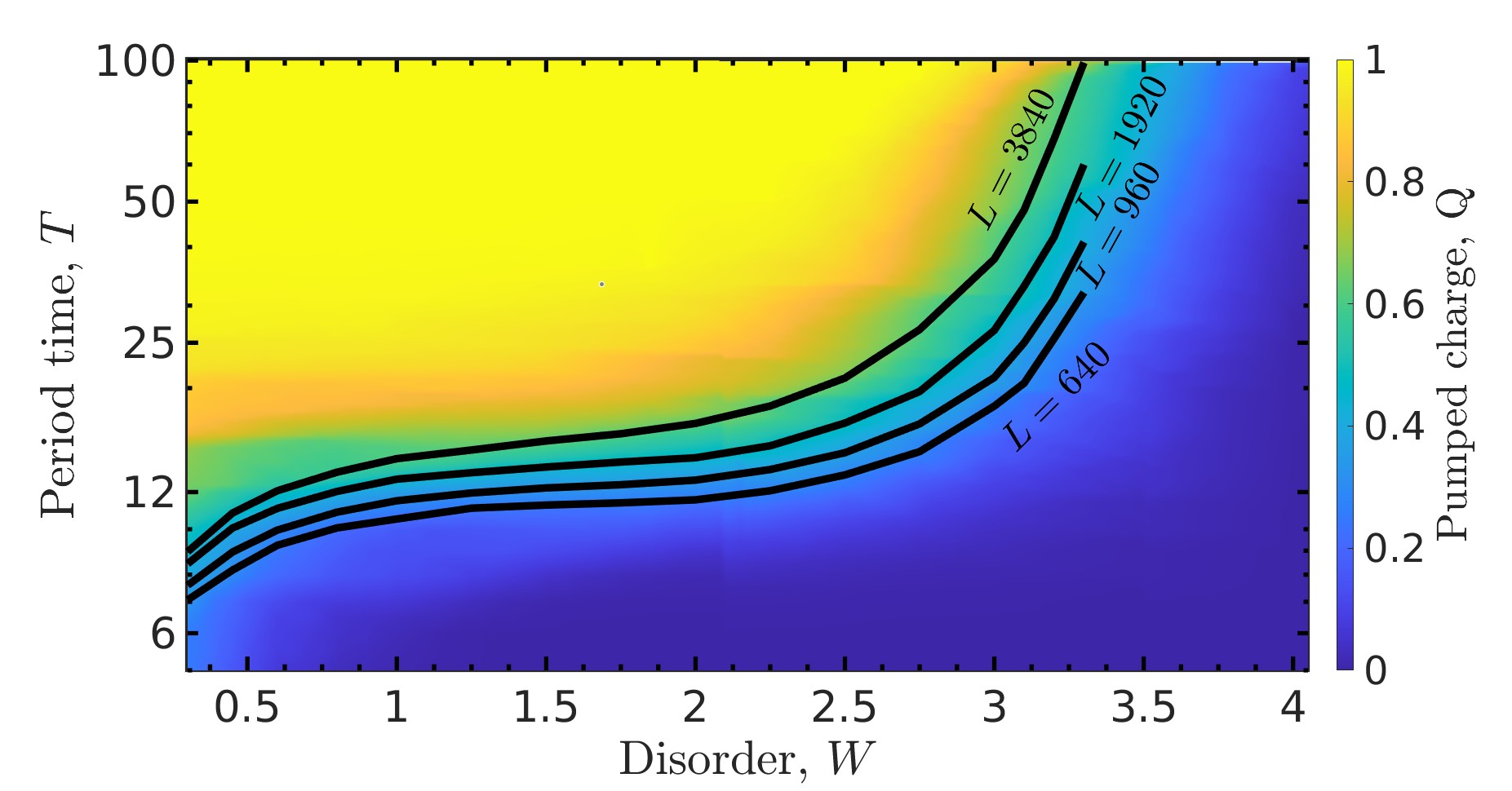}
\caption{
Colormap: Pumped charge $Q$, for $L=320$, for various period times $T$ and disorders $W$ (average over 20 disorder realizations).  
The Thouless pump works well (light area) for small $W$ and large $T$, and breaks down both if $T$ decreases or $W$ increases. 
For $W \gg W_c\approx3.5$, where there is no gap in the instantaneous energy spectrum, the pumped charge is very close to 0 for any $T$. 
Black lines: $T(W)$ curves along which $Q=1/4$. These show that the pump breaks down easier if the chain is longer.
}
\label{fig:map_T_W}
\end{figure}	

For a more complete picture of the breakdown of the Thouless pump as a function of disorder $W$, chain length $L$, and period time $T$, we show the numerically obtained map of pumped charge $Q$ in Fig.~\ref{fig:map_T_W}. 
The colors show $Q$ values for $L=320$, and results for other lengths are shown as $Q=1/4$ isolines.
These reveal four qualitatively different regimes of the charge pump. For small disorder, $W \lesssim 0.5$, 
the Floquet localization length $\zeta_F$ decreases sharply as the disorder is increased. 
For 
$0.5 \lesssim W \lesssim 2$, and period times $10 \lesssim T \lesssim 20$, $\zeta_F$ does not depend much on the disorder strength. 
For larger disorder, 
$2 \lesssim W \lesssim W_c = 3.5$, we have a sharp decrease of the $\zeta_F$ as $W$ is increased. 
Finally, above the critical disorder value, $3.5 \lesssim W$, we observe the charge pump breaking down completely.  

\begin{figure}
\includegraphics[width=\columnwidth,trim={0cm 0cm 0cm 0cm},clip]{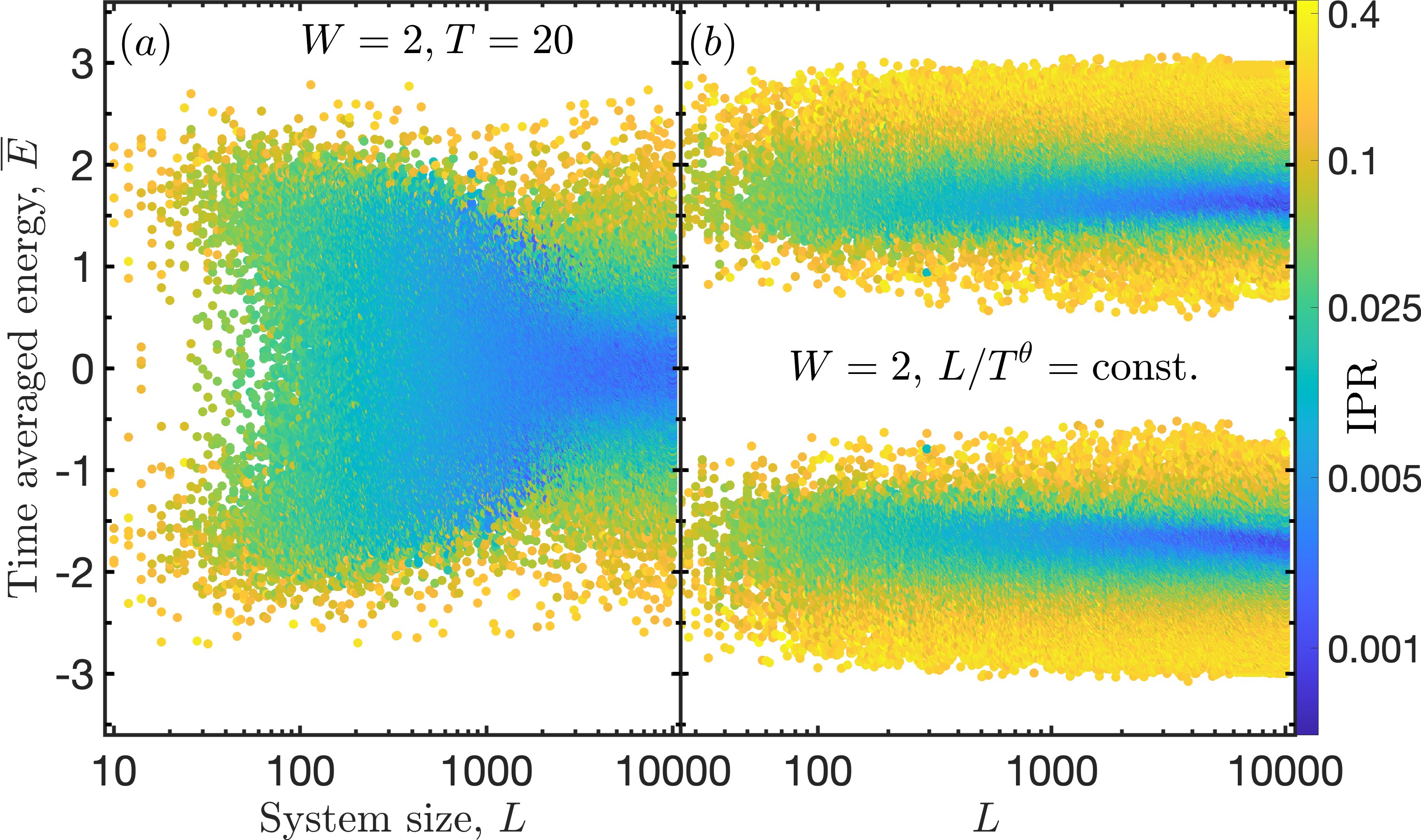}
\caption{
Spectrum of the average energy of Floquet states in a pump with disorder $W=2$, for various system lengths, and with color representing the IPR value of the states (single disorder realizations). 
In (a), the  period time fixed, $T=20$, increasing the length leads to a qualitative change in the spectrum: 
For $L\lesssim 100$, two bands of extended states can be seen, which carry current right/left in the lower/upper band; 
for larger lengths the gap between the bands is closed.
In (b), the period time is increased together with the system size, keeping $L/T^\theta$ constant, where $\theta={6.01}$ was used.   
Here there is no qualitative change in the spectrum as $L$ is increased, however, the range of extended states becomes narrower.}
\label{fig:spectrum_fix_T_and_running_T}
\end{figure}

\emph{Thermodynamic limit.} --- 
For a deeper understanding of how disorder impacts the Thouless pump, we 
define an alternate thermodynamic limit: 
$L\to\infty$ and $T\to\infty$ together, with $L/T^\theta$ kept constant.
This is needed, e.g., to explore the extended/localized nature of the Floquet states, using the inverse participation ratio \cite{wauters2019localization} (IPR) $P_2=\sum_{m=1}^L \left| \psi_{n,m} \right|^4$. 
To show how this limit avoids the problem of the Anderson localization of Floquet states, see Fig.~\ref{fig:spectrum_fix_T_and_running_T}.
We choose parameters so that for short lengths, $L \lesssim 100$, the Thouless pump works well, and Floquet states form two well separated bands. 
If the length is increased at fixed $T$ (panel a), the two bands merge and the pump begins to break down. 
In contrast, if $L/T^\theta$ is kept constant (panel b), the spectrum of Floquet states
shows no qualitative change up to the largest system sizes that we were able to access numerically. 
For each band, states in the band center are more extended (lower IPR, decreasing with $L$) and at the band edges more localized (higher IPR, independent of $L$).
More detailed analysis of the IPR values, and examples for shorter period times $T$, where the breakdown of the pump is even more visible, are shown in the SM \cite{SM}.

\emph{Discussion, conclusions.} ---
We found that onsite potential disorder in the periodically driven Rice-Mele model results in a breakdown of the Thouless pump in the $L\to\infty$, constant-$T$ limit, due to the Anderson localization of the Floquet states. 
This can be avoided by taking $L\to\infty$ and $T\to\infty$ together, keeping $L/T^\theta$ constant, where $\theta$ is a disorder-dependent critical exponent. 
Although we expected $\theta=2$ based on the corrections to adiabaticity, we found that this is not the case, rather, $\theta$ depends on disorder strength continuously. 
It is an interesting open problem to find an analytical explanation of this phenomenon.  

Our work is a starting point for a more systematic investigation of the relation between Anderson localization in the Thouless pump and the ``levitation and annihilation'' of extended states in Chern insulators. 
By finding a suitable thermodynamic limit, we open the way to studying numerically the conduction in the disordered Thouless pump, as well as the mechanism of its breakdown as disorder is increased. 
Our preliminary results, shown in the SM \cite{SM}, suggest that the current-carrying states here have a fractal nature, but it is an open question whether in the thermodynamic it is only a single Floquet state per band that carries the current.

From a broader perspective, the Thouless pump is the oldest among a large family of topological pumps, which by now go well beyond Chern insulators. 
Topological pumping has been associated with a wide range of topological insulators and superconductors \cite{Meidan2011, Ryu2012, Fulga2012}, and has been proposed to occur also between topological defects \cite{Teo2010, Xie2022}. 
It has been further extended to higher-order topological phases \cite{Benalcazar2017, Song2017, Schindler2018}, which can lead to dipole or to quadrupole pumps \cite{Benalcazar2017b}.
Very recently, topological pumping between the corners of a two-dimensional sample has been shown to occur both theoretically and experimentally, with the pump working either via bulk states \cite{Wienand2022, Benalcazar2022}, or via edge states \cite{Xia2023}.
Our work opens a new direction of research in this field, consisting in the study of the thermodynamic limits associated to this large family of pumps and the critical exponents characterizing them.

\emph{Acknowledgements}
We acknowledge support from the National Research, Development and Innovation Office (NKFIH) through the OTKA research grants No. K138606 and 132146, and within the Quantum National Laboratory of Hungary program (Grant No. 2022-2.1.1-NL-2022-00004).
ICF acknowledges support from the Deutsche Forschungsgemeinschaft (DFG, German Research Foundation) under Germany's Excellence Strategy through the W\"{u}rzburg-Dresden Cluster of Excellence on Complexity and Topology in Quantum Matter -- \emph{ct.qmat} (EXC 2147, project-ids 390858490 and 392019).

\bibliography{main.bib}

\end{document}


\title{Supplemental Material to ``Floquet-Anderson localization in the Thouless pump and how to avoid it''}

\author{Andr\'as Grabarits}
\affiliation{Department of Theoretical Physics, Institute of Physics, 
Budapest University of Technology and Economics, M\H uegyetem rkp. 3., H-1111 Budapest, Hungary}
\affiliation{MTA-BME Quantum Dynamics and Correlations Research Group, 
Budapest University of Technology and Economics,  M\H uegyetem rkp. 3., H-1111 Budapest, Hungary}
\author{Attila Tak\'acs}
\affiliation{Department of Theoretical Physics, Institute of Physics, 
Budapest University of Technology and Economics, M\H uegyetem rkp. 3., H-1111 Budapest, Hungary}
\affiliation{Université de Lorraine, CNRS, LPCT, F-54000 Nancy, France}
\author{Ion Cosma Fulga}
\affiliation{Leibniz Institute for Solid State and Materials Research,
IFW Dresden, Helmholtzstrasse 20, 01069 Dresden, Germany}
\affiliation{W\"{u}rzburg-Dresden Cluster of Excellence ct.qmat, 01062 Dresden, Germany}

\author{J\'anos K. Asb\'oth}
\affiliation{Department of Theoretical Physics, Institute of Physics, 
Budapest University of Technology and Economics, M\H uegyetem rkp. 3., H-1111 Budapest, Hungary}
\affiliation{Wigner Research Centre for Physics, H-1525 Budapest, P.O. Box 49., Hungary}

\date{\today}

\begin{abstract}
In this Supplemental Material, we provide more details on the calculation and scaling of the pumped charge, as well as on the Anderson localization of the Floquet states.
\end{abstract}

\maketitle

\section{Numerical calculation of the exponent $\theta$}

We provide more details of the numerics behind the result shown in Fig.~2.~of the main text: the critical exponent $\theta$ does not have a universal value, rather depends continuously on the disorder $W$.
We use the scaling Ansatz also included in the main text,  
\begin{align}
\label{eq:main_scaling_exponent_def}
    Q(L, T, W) &= Q \left( (T'/T)^{\theta(W)} L,\, T',\, W \right) ,
\end{align}
whereby for a fixed disorder strength $W$, the pumped charge $Q$ is invariant to changing the period time from $T$ to any $T'$, as long as we also change the chain length to $L'$, with $L' =(T'/T)^{\theta(W)} L$. 
We used three different numerical methods to obtain this result, as detailed below. 

\subsection{Scaling at fixed $Q=1/2$, $Q=1/4$}

The first two approaches to obtain $\theta$ used the scaling relation Eq.~\eqref{eq:main_scaling_exponent_def}, and attempted the best possible collapse at $Q=1/2$ and at $Q=1/4$, respectively. 
Thus, at each fixed value of disorder $W$, for system sizes varying from $L=80$ to $L=2240$ we selected $5-6$ such $T$ values for which the calculated pumped charges
of $20$ disorder realizations were within the range of $\Delta Q=0.1$ around the predefined $Q=1/4,\,1/2$ values. We then determined the value of $T$ of $Q=1/4$ or $Q=1/2$ by performing a linear regression on these data. In addition we also checked that the value of the obtained period times converged properly with the number of charge values involved in the linear regression, meaning that involving more charge values in the linear regression corresponding either to the used period times or to new period times does not change the obtained period times relevantly. We also ruled out finite-size effects by recalculating $\theta$ with the data points $(L, T)$ with the smallest system sizes excluded, and found no significant changes. 

The calculation with $Q=1/4$ is actually numerically more efficient than that with $Q=1/2$. The reason is that for the same chain length $L$, lower values of the  pumped charge require faster pumps, smaller values of $T$, and hence, fewer time-slices for the calculation of the Floquet states. Thus, while with $Q=1/2$, we could only reliably extract $\theta$ for disorder strengths $W \gtrsim 1$, with $Q=1/4$, we could compute $\theta$ for $W \gtrsim 0.3$. These features can be seen in Fig.~\ref{fig:Fittings}.

\subsection{Three-parameter fit}

\begin{figure}
\includegraphics[width=\textwidth]{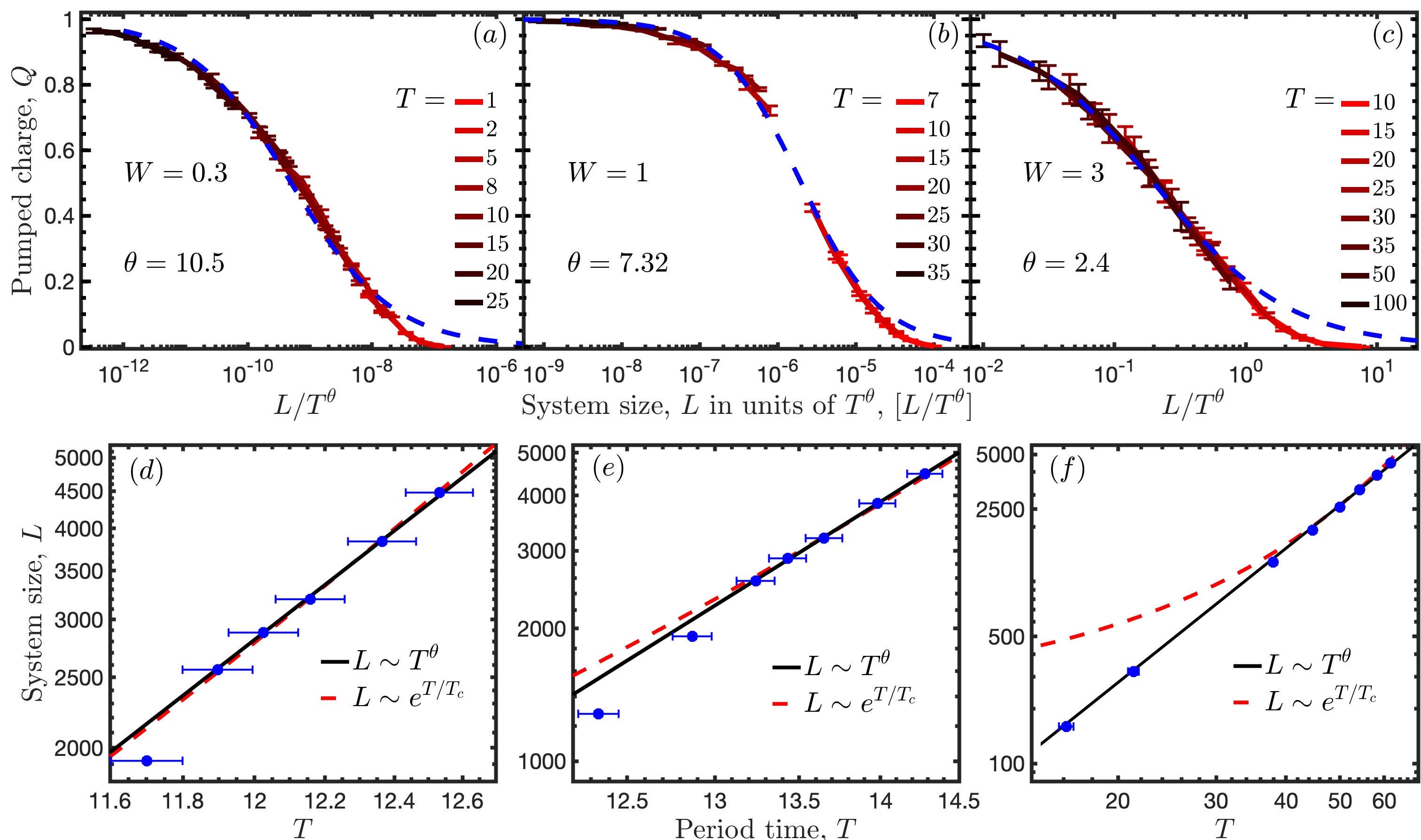}
\caption{Numerics for testing the scaling collapse of the pumped charge $Q$ and the extraction of $\theta$. Panels (a-c): $Q$ as a function of $L/T^\theta$, together with the fitted Ansatz of Eq.~\eqref{eq:Qfit}, for disorder $W=0.3$, $1$, and $3$, respectively. Panels (d-f): The fitting of the numerically obtained $T-L$ pairs where the charge is around (d): $Q=1/4$, for $W=0.3$; (e): $Q=1/2$, for $W=1$; (f): $Q=1/2$, for $W=3$. Besides the power-law Ansatz of Eq.~\eqref{eq:Qfit}, logarithmic scaling Ansatz of Eq.~\eqref{eq:exponential_scaling} is also used to fit the data. The former approach provides much more precise fitting for $W=3$, is only a little bit better for $W=1$, while for $W=0.3$ the two fit functions provide indistinguishably good fits to the results.}
\label{fig:Fittings}
\end{figure}

Our third numerical approach was, at each fixed value of disorder $W$, to fit an Ansatz to all of the numerical values of $Q(W, L, T)$. 
We used a three-parameter Ansatz, related to the exponent $\theta$, 
\begin{align}
\label{eq:Qfit}
Q_\mathrm{fit}(L, T, W) &=  \frac{1}{2}-\frac{1}{2} 
\tanh [ a_1(W) \ln L + a_2(W) \ln T + a_3(W) ];&
\theta(W) &= - \frac{a_2(W)}{a_1(W)}.
\end{align}
We chose this Ansatz because it could be fitted quite well to the data in the range of parameters where $0.9>Q>0.2$ --  when the pump works reasonably well, but corrections to the quantized value of the pumped charge are already sizable --  as seen in Fig.~\ref{fig:Fittings}. 
We used chain lengths $L=80, \ldots, 1280$ and varied the period time in the range $T \in[1,200]$ providing charge values ranging from $0$ to $1$ for all disorder strengths. Here the pumped charge values were obtained by averaging over $20$ disorder realizations and similarly to the $Q=1/4,\,Q=1/2$ approach we checked that the these average values did not change relevantly when increasing further the number of disorder realizations.
In addition, we also checked that the obtained exponents did not change significantly when excluding the smallest system sizes from the fit.

We show examples for the fits for small, intermediate, and large disorder  strengths, $W=0.3$, $1$, and $3$, in Fig.~\ref{fig:Fittings}. The charge values show a good collapse as a function of $L/T^\theta$ for different system sizes. 

\subsection{Small disorder: data range too small for reliable demonstration of power-law scaling} \label{susec:alternative_scaling}

At small disorder values, $W < 1$, increasing the chain length $L$ by a factor of 30, we can keep the pumped charge $Q$ the same by taking slightly longer period time $T$ (less than 60\% increase). This is consistent with the power-law scaling of  Eq.~\eqref{eq:main_scaling_exponent_def}, with exponent $\theta$ that can fitted precisely (and turns out to be $\theta>7$). However, this range of $T$ is not enough to provide conclusive evidence for the power-law scaling in this disorder regime. Spanning a larger range of system sizes is beyond our computational capacity. 

\begin{figure}
\includegraphics[width=0.7\textwidth]{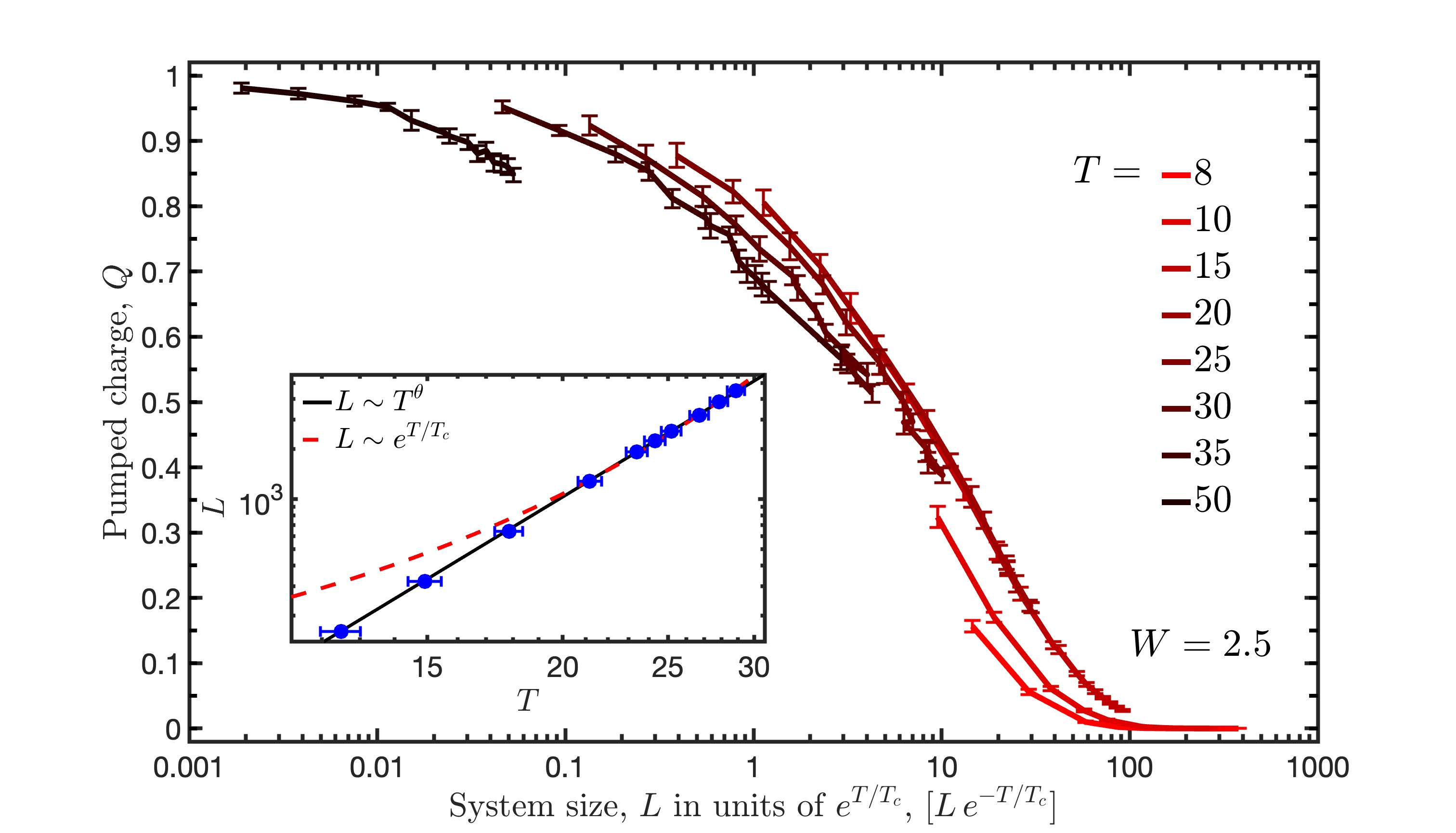}
\caption{Attempt at collapsing the numerical results for the pumped charge $Q$ at disorder $W=2.5$, using the logarithmic Ansatz of Eq.~\eqref{eq:exponential_scaling}, with $T_c = 6.14$ giving the best fit. The collapse is visibly less precise than with the power-law scaling of Eq.~\eqref{eq:main_scaling_exponent_def} in the main text with $\theta=3.95$. Inset: the $(L,T)$ pairs where $Q\approx 1/4$ clearly fall on a straight line on the log-log scale, and are better fitted with the power-law Ansatz than by the logarithmic Ansatz.}
\label{fig:APPexpbeta}
\end{figure}

One possible alternative scaling law, compatible with our data at small disorders, is: 
\begin{align}
\label{eq:exponential_scaling}
    Q(L, T, W) = Q \left( L', T + T_c(W) \ln(L'/L), W \right) .
\end{align}
At disorders $W=2$ and above, however, this alternative scaling law is fits the data much worse than that of Eq.~\eqref{eq:main_scaling_exponent_def}. We show 
this in Fig.~\ref{fig:APPexpbeta}. It can clearly be observed that for large enough disorders the power-law $L\sim T^\theta$ provides much more reliable fitting, further enforcing our assumption about structure of the correct thermodynamic limit.

\section{Thouless pump without disorder: details of computation}

In this section, we provide details about how, for the Thouless pump without disorder, $W=0$, we calculated the pumped charge $Q$ for large system sizes, $L\sim 10^6$. 
We followed the approach of Privitera et al.~\cite{Privitera2018}.

Without disorder, we have translation invariance, and thus the pumped charge can be calculated separately for each quasimomentum $k$. Thus, we only need to calculate with matrices of size $2\times 2$.   
Using $\omega=2\pi/T$, the Hamiltonian matrix reads, 
\begin{equation}
 \mathcal H_k(t)=\begin{bmatrix}
 \Delta \sin\omega t & J(1+e^{ik}) + \tilde J (1-e^{ik}) 
 \cos\omega t \\
 J(1+e^{-ik}) + \tilde J  (1-e^{-ik}) \cos \omega t &-\Delta \sin\omega t
 \end{bmatrix}.
\end{equation}
Its eigenstates are 2-component vectors $\ket{\phi_+(k,t)}$ and $\ket{\phi_-(k,t)}$, with the corresponding energies, $\pm E(k,t)$, with 
\begin{equation}
    E(k,t)^2 = 
    2 J^2 + 2 \tilde J^2 \cos^2\omega t
    + \Delta^2\sin^2\omega t 
    + 2 (J^2-\tilde J^2 \cos^2\omega t ) \cos k .
\end{equation}
The unitary timestep operator can be obtained by multiplying the operators of the time-slices,
\begin{align}
U_k &= U_k(T) U_k(T-\mathrm d t) \ldots U_k(\mathrm d t);&
U_k(t) &= e^{-i\mathrm{d}t \mathcal H_k(t)} = 
 \cos[E(k, t)\mathrm dt]
 -i \frac{\sin(E(k, t)\mathrm dt)}{E(k, t)} 
 \mathcal H_k(t).
\end{align}
The eigenstates of $U_k$ are the Floquet states, 2-component vectors $\ket{\psi_1(k)}$ and $\ket{\psi_2(k)}$ with the quasienergies $\varepsilon_{1,2}(k)$ (unique up to mod ${2\pi}/{T}$) defined by 
$
    U_k\ket{\psi_{1,2}(k)}=e^{-i\,\varepsilon_{1,2}(k)T}\ket{\psi_{1,2}(k)}.
$

The pumped charge $Q$, in the limit of sustained pumping \cite{Privitera2018}, and its discretized version, read
\begin{equation}
	Q
 =\sum_{n=1}^2\int_{-\pi}^{\pi}\mathrm dk\frac{\partial \varepsilon_n(k)}{\partial k}
\left| \braket{\phi_{-}(k,0)|\psi_{n}(k)} \right|^2.
\approx 
\frac{i}{\Delta k} \sum_{n=1}^2 \left[ \sum_{l=2}^{L/2} \ln\left( \frac{e^{-i\varepsilon_n(k_l)}}{e^{-i\varepsilon_n(k_l-\Delta k)}} \right) \left| 
\braket{\phi_{-}(k_l,0)|\psi_{n}(k_l)} \right|^2 \right], 
\end{equation}
with $\Delta k=\frac{2\pi}{L/2}$ and $k_l=-\pi+l\frac{2\pi}{L/2},\,l=1,\,\dots,\,L/2$.

\section{Spectrally resolved analysis of Floquet states for disorder $W=2$ in two competing approaches to a thermodynamic limit}

We provide some more detailed analysis of the Floquet spectra of the disordered Thouless pump, including the inverse participation ratio (IPR) values of the Floquet states, for disorder $W=2$, as shown in Fig.~4 of the main text. The IPR of the $n$th Floquet state is defined as $P_2=\sum_{m=1}^L \left| \psi_{n,m} \right|^4$, where $\psi_{n,m}$ is the probability amplitude of site $m$ in the Floquet state $\ket{\psi_n}$.
    	
\begin{figure}
\includegraphics[width=0.49\textwidth]{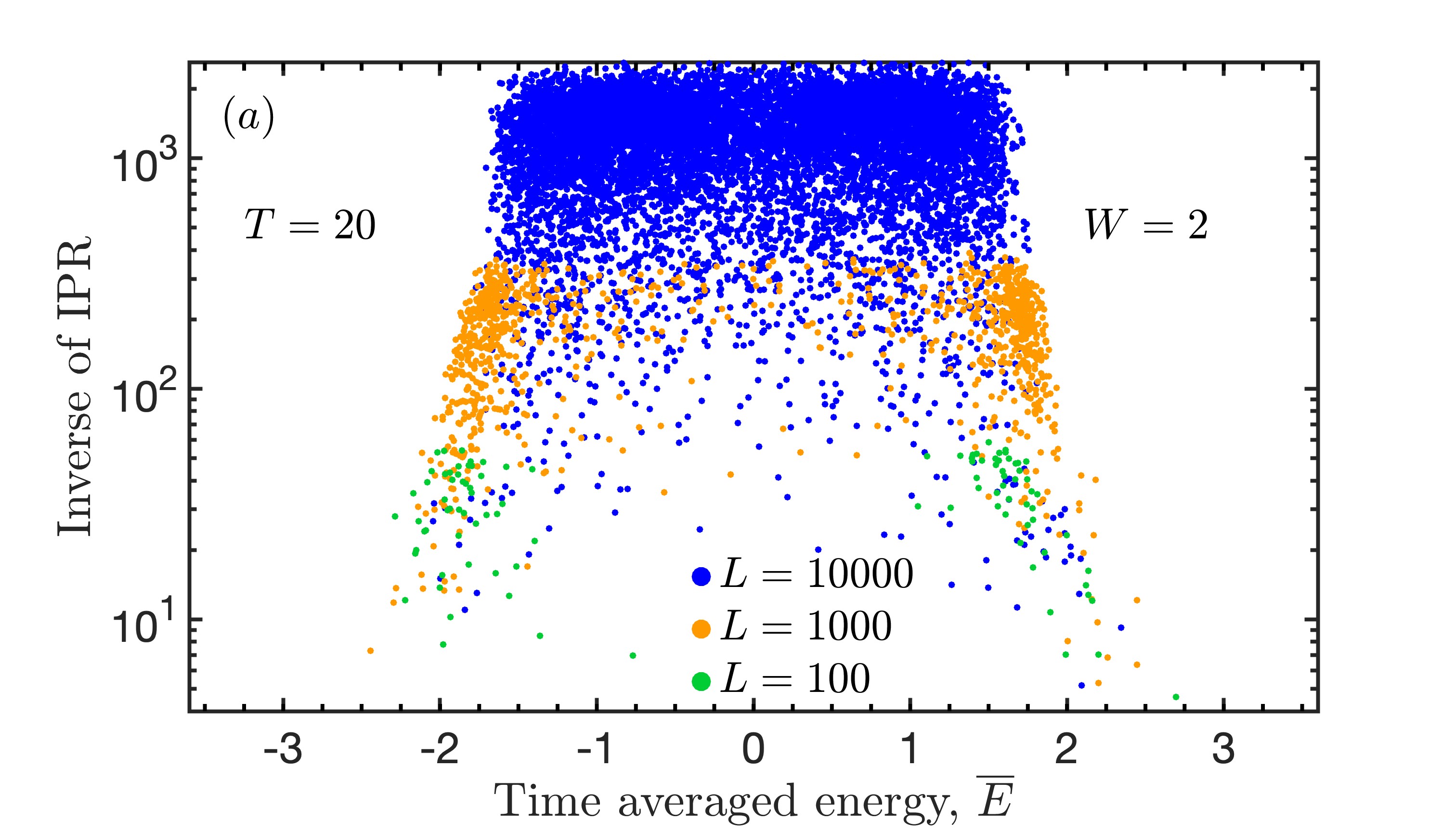}
\includegraphics[width=0.49\textwidth]{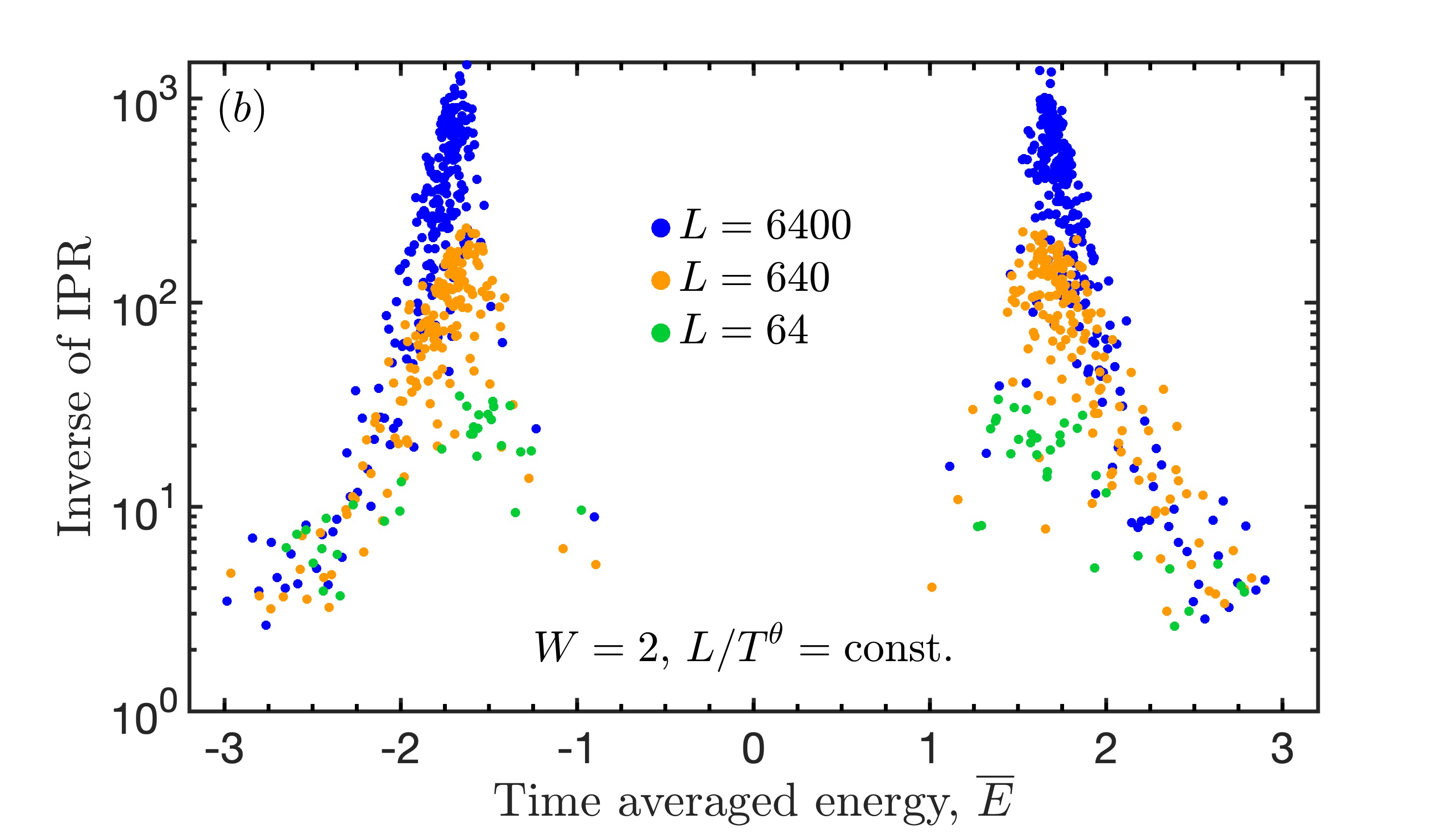}
\caption{Inverse IPR values of Floquet states, with disorder $W=2$, for various system sizes $L$, with (a) fixed $T=20$ and (b) fixed $L/T^\theta=3.4\times10^{-10}$.  
In both cases, the inverse IPR values of the most delocalized states increase as the system size increases.  However, in (a) the average energy gap closes as the system size increases, while in (b) the gap remains open, and we see states with long localization lengths in the middle of the bands.}
\label{fig:IPR_spectra_as_main_text}
\end{figure}

\begin{figure}
\includegraphics[width=0.49\textwidth]
{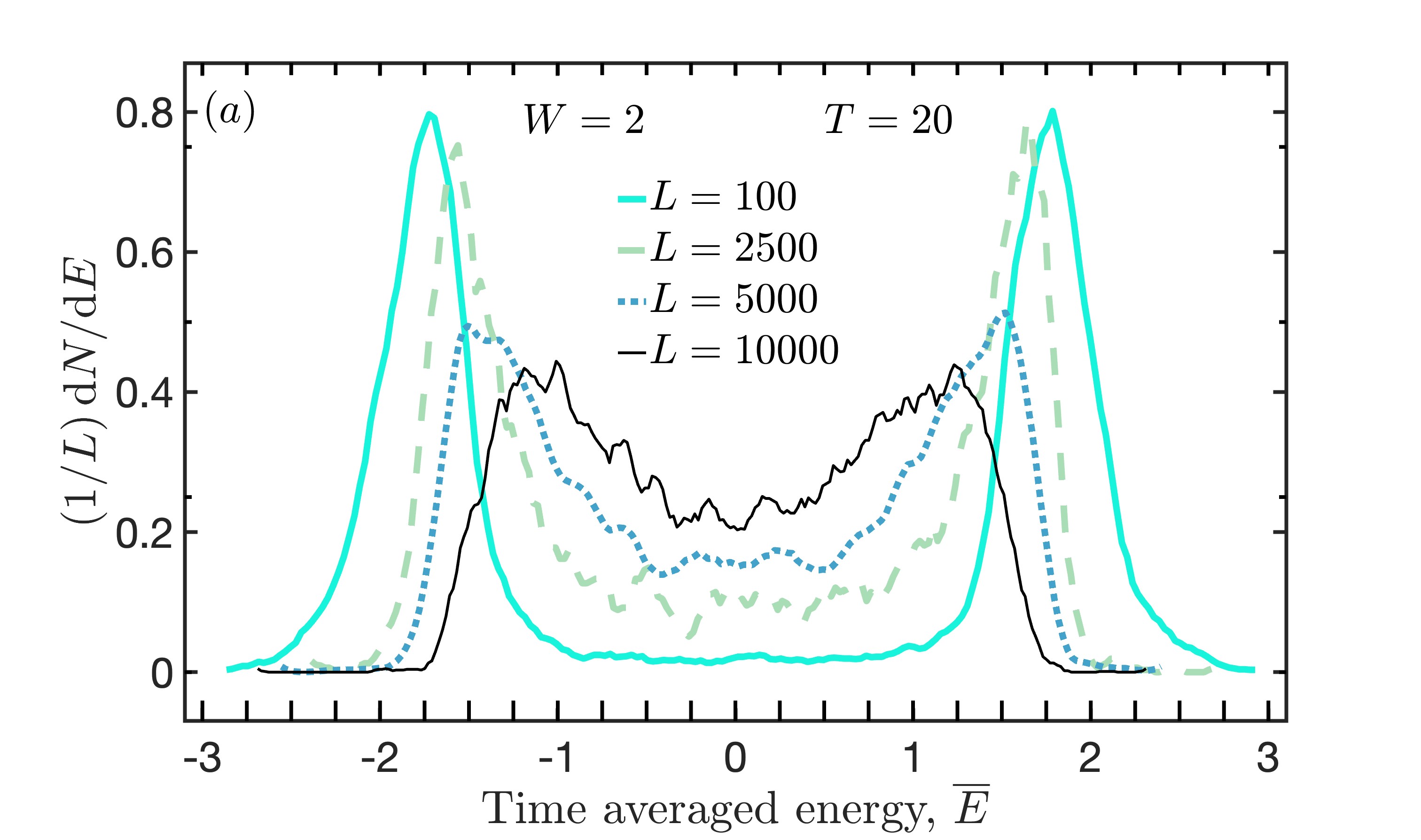}
\includegraphics[width=0.49\textwidth]
{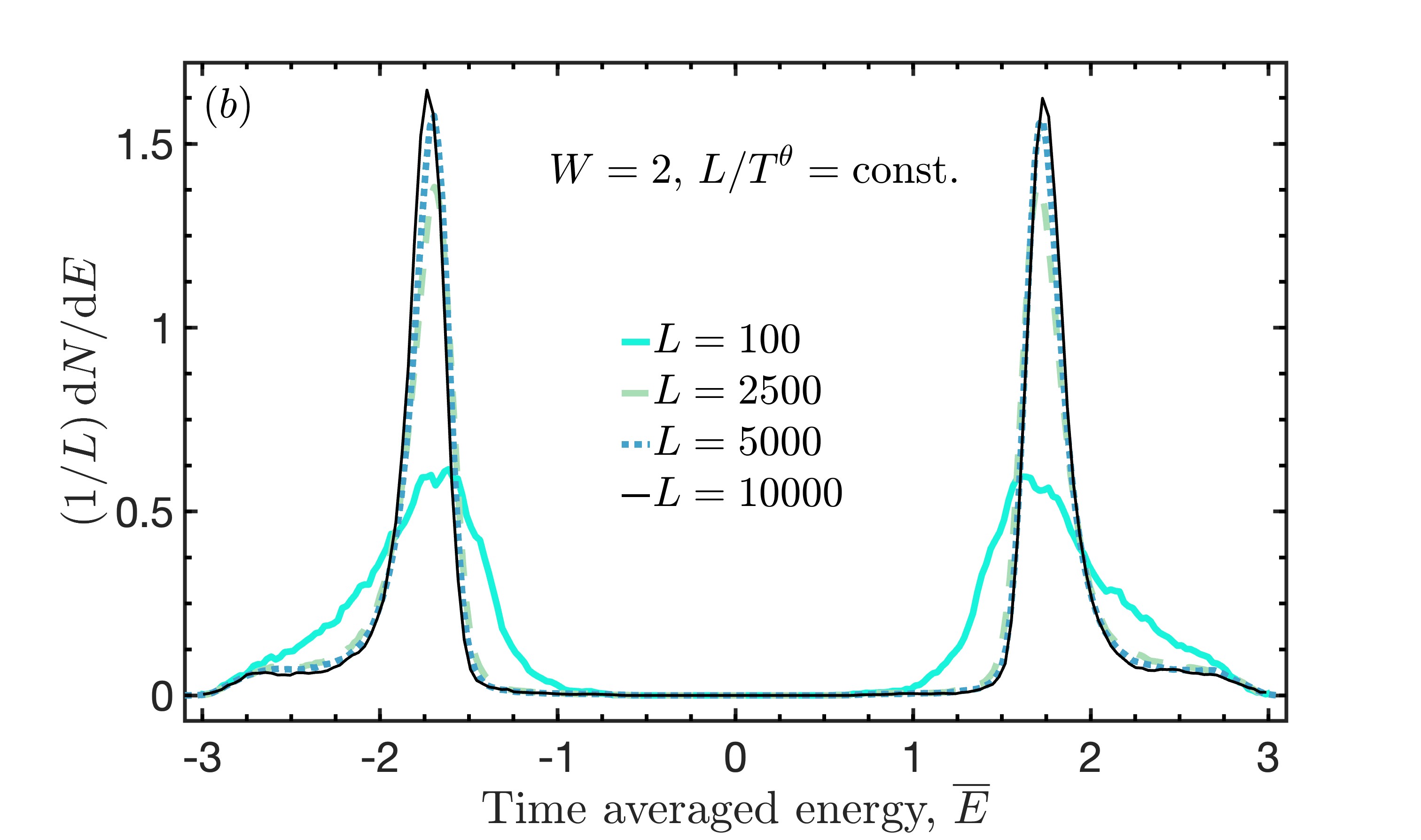}
\caption{ Spectrally resolved density of states per unit length, at disorder $W=2$, for increasing system sizes. Panel (a): with fixed $T=20$, increasing the system size leads to a merging of the bands.  Panel (b): increasing $T$ as well, keeping $L/T^\theta=3.4\times10^{-10}$, we see well separated bands, whose density of states curves fall on top of each other for the largest system sizes, indicating a well-defined thermodynamic limit. }
\label{fig:DOS}
\end{figure}

\begin{figure}
\includegraphics[width=0.49\textwidth]{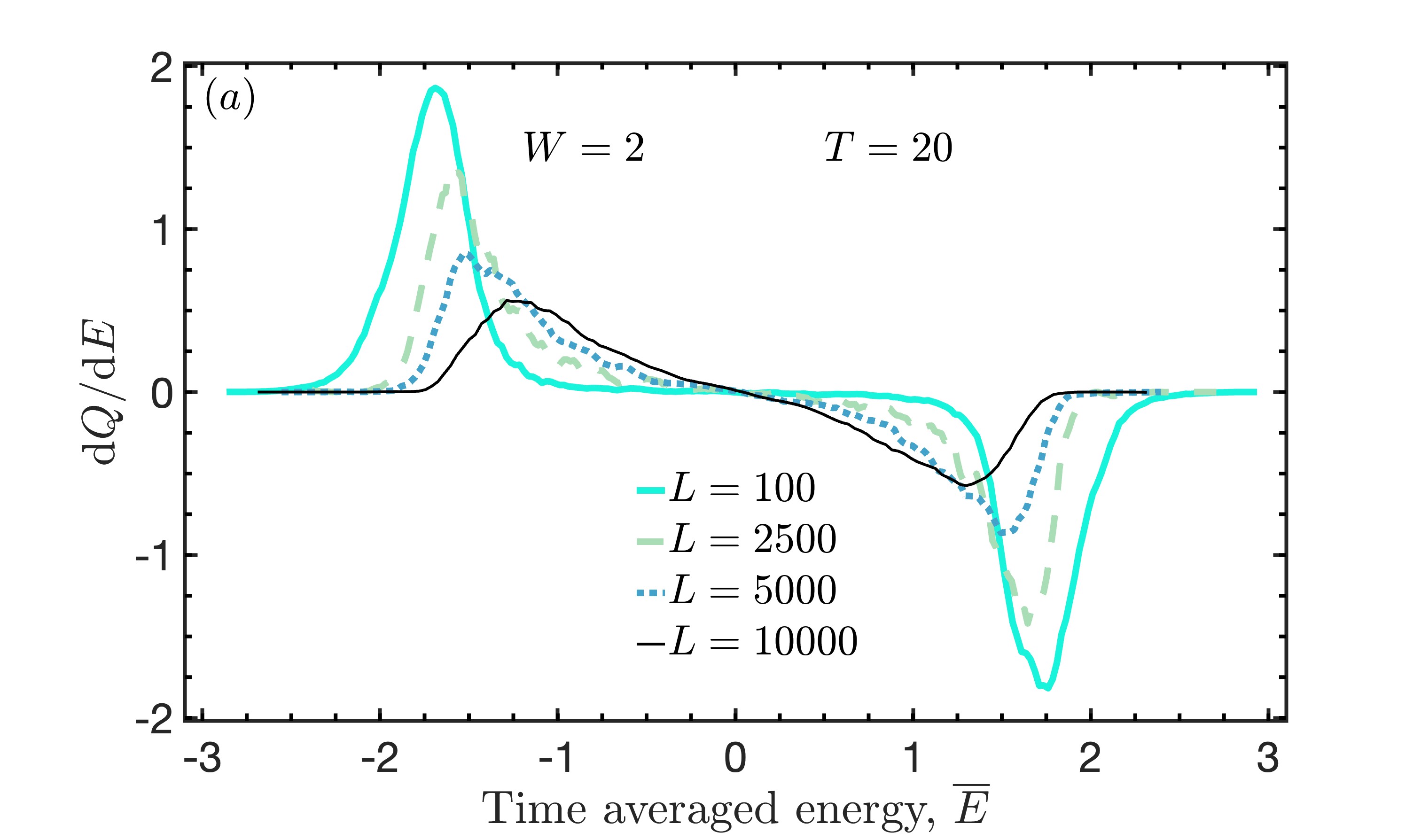}
\includegraphics[width=0.49\textwidth]{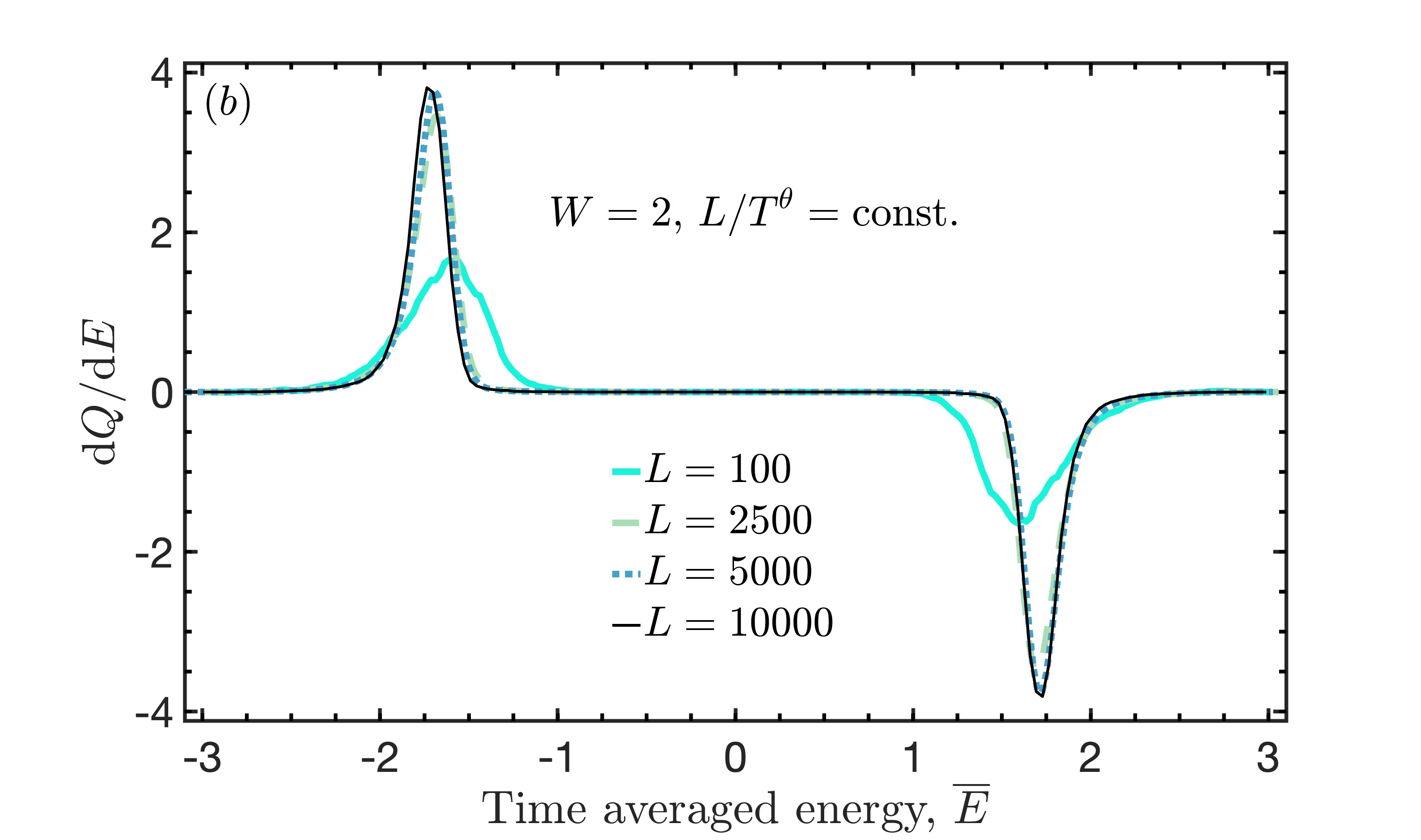}
\caption{Spectrally resolved pumped charge at disorder $W=2$, for increasing system sizes. Panel (a): with fixed $T=20$, for the larger system sizes there is a wider range of current-carrying Floquet states, but the total current carried by states with negative average energy is significantly less than 1, $Q \approx 0.22$?   
Panel (b): keeping $L/T^\theta=3.4\times10^{-10}$,$L/T^\theta$ constant, the curves fall on top of each other for the largest system sizes, indicating a well-defined thermodynamic limit, with pumped charge $Q\approx0.9996$.
\label{fig:dQdE}}
\end{figure}

First, in Fig.~\ref{fig:IPR_spectra_as_main_text}, we show some more detailed plots of properties of Floquet states at disorder $W=2$, data which was partly used for Fig.~4 of the main text.  
Fig.~\ref{fig:IPR_spectra_as_main_text} shows the inverse of the IPR values, $1/P_2$, 
of the Floquet states $\ket{\psi_n}$ for some system sizes, 10 random realizations for each sets of parameters. 
The two subplots correspond to the subplots of Fig.~4 of the main text, i.e., scaling up the system size $L$, while (a) keeping the period time constant, or (b) keeping $L/T^\theta$ constant. In the first case, we see a gradual merging of the two bands, while in the second, the two bands remain well separated in average energy even at the largest system sizes.  In both (a) and (b), however, the 1/IPR values increase as the system size is increased, roughly as $P_2 \propto L^{-0.75}$. A more detailed investigation of this scaling is work in progress.

Some more details extracted from these simulations are shown in Figs.~\ref{fig:DOS} and \ref{fig:dQdE}: the spectrally resolved density of states per unit length, and spectrally resolved pumped charge ($\mathrm dQ/\mathrm dE$). On both pairs of plots, if the size is increased while keeping $T$ fixed (panels a), we see early signatures of Anderson localization (merging of the bands and a decrease of the current in the middle of the spectrum). If we keep $L/T^\theta$ constant, however (panels b), we see two separated bands that carry current, and an approach to a thermodynamic limit, where the curves for $L=5000$ and $10000$ fall on top of each other.  

\begin{figure}[h]
\includegraphics[width=0.49\textwidth]{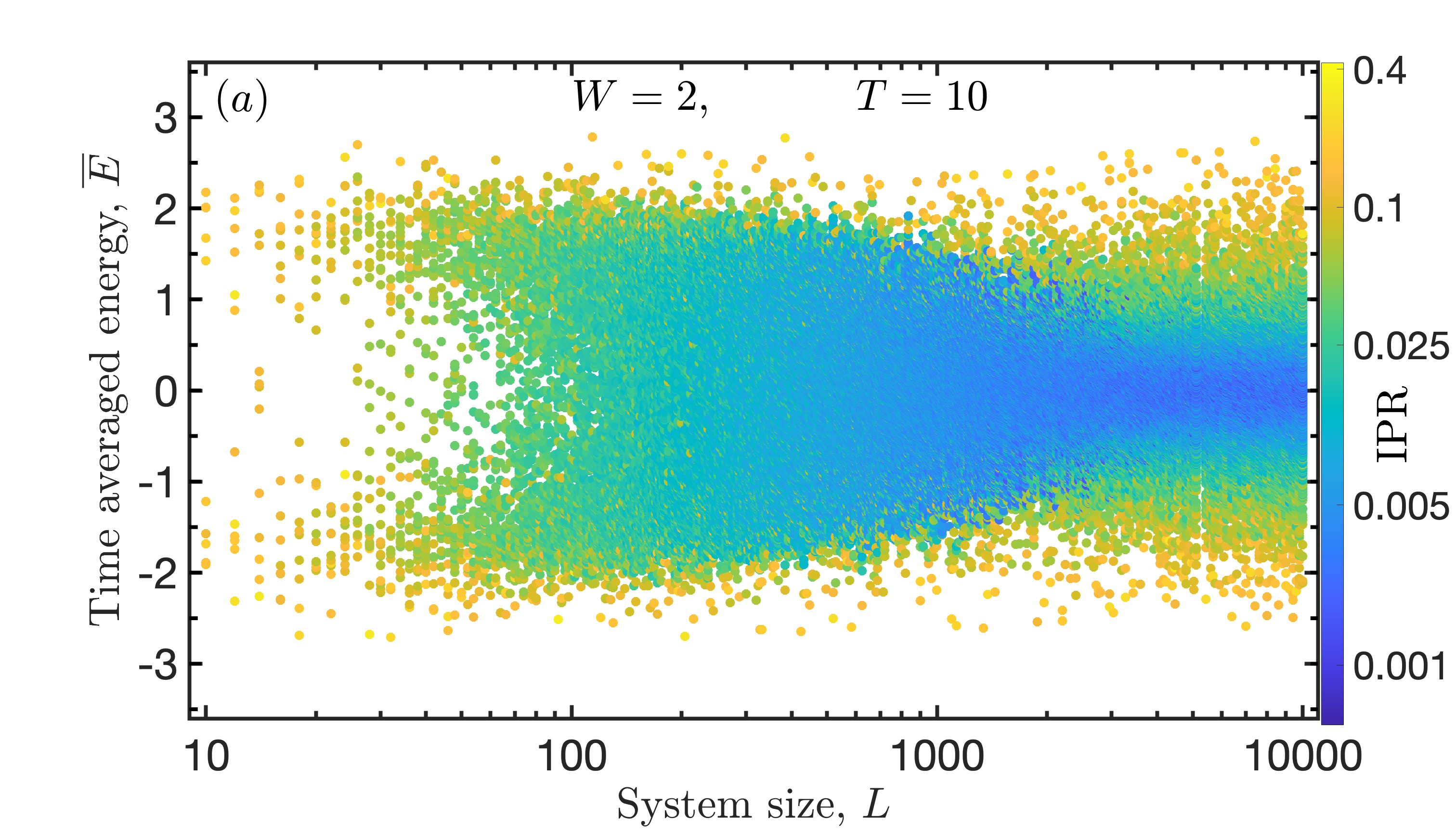}
\includegraphics[width=0.49\textwidth]
{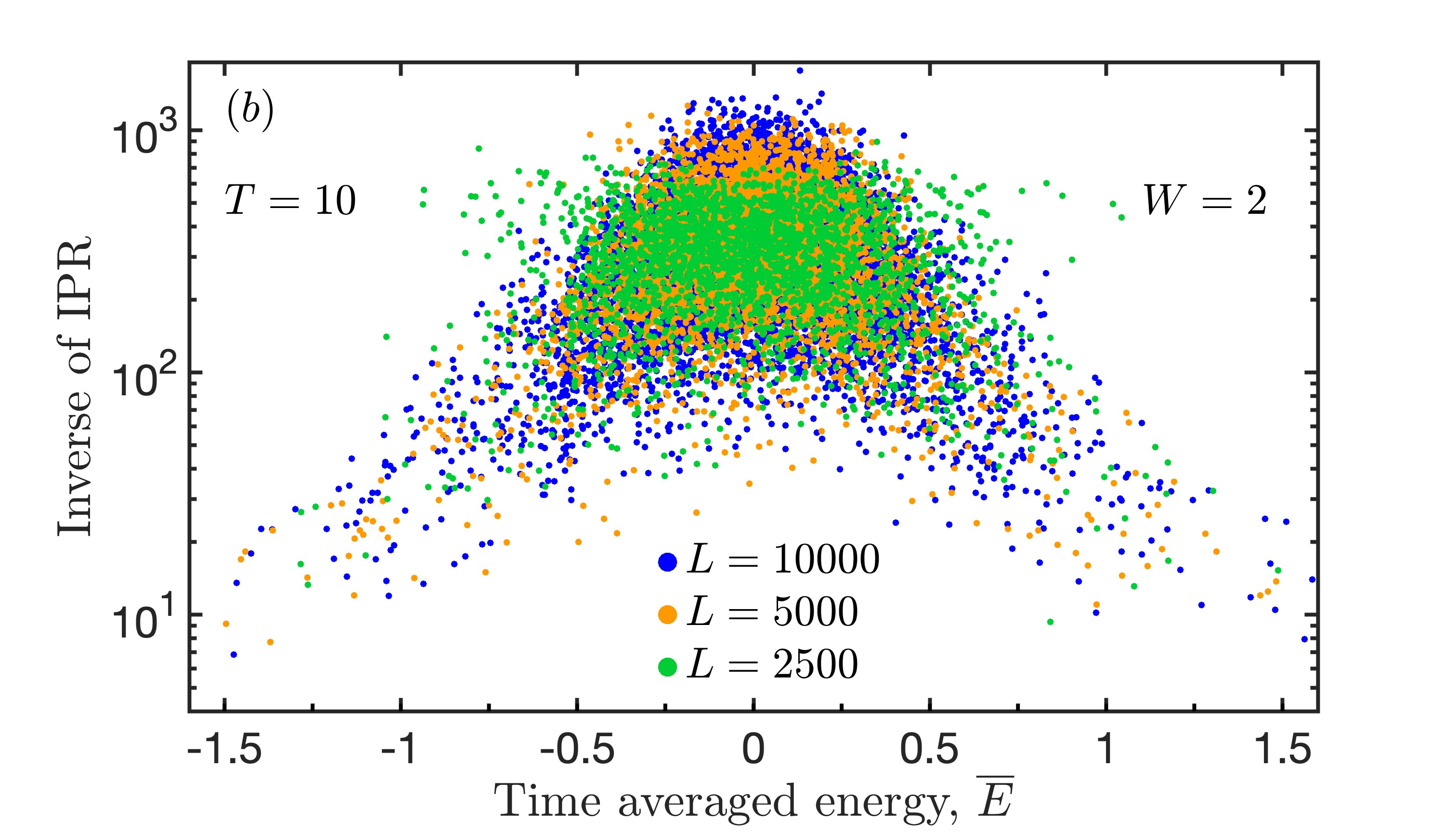}
\caption{Panel (a):  Time-averaged energy of Floquet states for $W=2$ for system sizes $L=10, \ldots, 10000$ and for fixed period time $T=10$. Color indicates IPR of the Floquet states. 
Panel (b): Cuts of the map plot of panel (a) for system sizes $L=2500, 5000$, and $L=10000$ (with inverse IPR displayed). 
For small system sizes two energy bands can be observed. Around $L\approx 100$  the gap between the bands closes. Then the density of states becomes more concentrated around $\overline{E}=0$. For even larger systems, around $L \gtrsim 2500$, both the density of states and the IPR values appear to not change much as $L$ is further increaesed.
\label{fig:Floquet_spectra_T_10}
}
\end{figure}

We also show data from other numerical simulations at disorder $W=2$, with a shorter period time, $T=10$, in Fig.~\ref{fig:Floquet_spectra_T_10}. Here the evidence of Anderson localization is more clear. In panel (a), just as in Fig. 4 of the main text, each point corresponds to a Floquet state, with color representing the IPR value, while in panel (b), just as in Fig.~\ref{fig:IPR_spectra_as_main_text} we show the inverse of the IPR values for selected system sizes. Both panels not only show that the two energy bands merge as the system size is increased, but also but also that the IPR values saturate, with $1/P_2$ essentially unchanged when $L$ is increased if $L \gtrsim 2500$.

\bibliography{main.bib}